\documentclass{article}
\usepackage{spconf,amsmath,graphicx}
\usepackage{CJKutf8}
\usepackage{multirow}
\usepackage{multicol}
\usepackage{booktabs}
\usepackage{hyperref}
\usepackage{cite}

\title{Listen and Speak Fairly: A Study on Semantic Gender Bias in Speech Integrated Large Language Models}
\name{\parbox{0.8\linewidth}{\centering Yi-Cheng Lin, Tzu-Quan Lin$^*$, Chih-Kai Yang$^*$, Ke-Han Lu$^*$, Wei-Chih Chen$^*$, Chun-Yi Kuan$^*$, Hung-yi Lee}}
\address{National Taiwan University, Taiwan}
\begin{document}
\ninept
\maketitle
\def\thefootnote{*}\footnotetext{Equal second contribution}\def\thefootnote{\arabic{footnote}}
\begin{abstract}
Speech Integrated Large Language Models (SILLMs) combine large language models with speech perception to perform diverse tasks, such as emotion recognition to speaker verification, demonstrating universal audio understanding capability. However, these models may amplify biases present in training data, potentially leading to biased access to information for marginalized groups. This work introduces a curated spoken bias evaluation toolkit and corresponding dataset. We evaluate gender bias in SILLMs across four semantic-related tasks: speech-to-text translation (STT), spoken coreference resolution (SCR), spoken sentence continuation (SSC), and spoken question answering (SQA). Our analysis reveals that bias levels are language-dependent and vary with different evaluation methods. Our findings emphasize the necessity of employing multiple approaches to comprehensively assess biases in SILLMs, providing insights for developing fairer SILLM systems.
\end{abstract}
\vspace{-5pt}
\section{Introduction}
\label{sec:intro}
Large Language Models (LLMs) have shown promising capability in recent years. However, LLMs cannot perceive non-textual modalities such as audio or image like humans.
Speech is an important modality in human communications, providing more information such as emotion and speaker than text cannot.
Many previous works try to incorporate LLMs with speech perception capabilities, resulting in Speech-Integrated Large Language Models (SILLMs). These models enable tasks such as speech translation and speech question answering (SQA) \cite{qwen_audio, ltuas, audiochatllama, tang2024salmonn}. 

SILLMs include two main types: cascaded systems, which combine automatic speech recognition (ASR) with LLMs, and end-to-end systems known as speech large language models (SLLMs). Cascaded systems use an ASR model to transcribe speech into text, which is then processed by an LLM. In contrast, SLLMs are designed to handle speech inputs directly and perform end-to-end processing.

SILLMs use robust LLMs pre-trained on large-scale internet data like Wikipedia and Reddit and fine-tuned on extensive speech-text paired data.  
However, imbalances in the domain, type, and speaker representation in the training data can introduce and amplify biases, leading to stereotypical outputs \cite{yeh2023evaluating, liang2021towards, hutchinson2020social}. 
For instance, 87\% of the editors on Wikipedia are male \cite{glott2010wikipedia}, and less than 19\% of English biographies are about women \cite{tripodi2023ms}, which might lead to a skewed representation of females in such a widely-used dataset \cite{kotek2023gender}, causing unintended consequences in deployed SILLM.

Biased SILLM may affect the society in several aspects. In education, students relying on SILLM for speech-to-text translation to learn languages and access educational content may receive biased or incorrect information, impacting their learning outcomes \cite{merine2022risks, sha2021assessing}. In business, biased SILLM creates barriers for non-native speakers and marginalized groups in accessing information \cite{1579399}, services \cite{malik2023linguistic}, and opportunities \cite{russo2017non}, impacting economic mobility and inclusion \cite{agiza2024analyzing}. However, despite there are some works on analyzing the stereotype and bias in LLM \cite{bbq, crows_pairs, wan2023kelly} and Multi-modal LLM \cite{hendricks2018women, seth2023dear, shankar2017no}, none of them discuss speech-integrated LLMs.

This study introduces a toolkit and the corresponding dataset for evaluating gender bias in SILLMs across four semantic-related tasks: speech-to-text translation (STT), spoken coreference resolution (SCR), spoken sentence continuation (SSC), and spoken question and answering (SQA). 

In STT, translating from English to grammatically gendered languages is challenging because these languages have complex gender systems that require nouns, articles, verbs, and adjectives to match the noun's gender. This can cause incorrect translations from English to languages like Spanish, due to English's lack of explicit gender indicators \cite{winomt}. 
In SCR, a model may exhibit coreference bias when it links gendered pronouns to occupations stereotypically associated with that gender more accurately than to occupations not associated with that gender \cite{winobias}.
In SSC and SQA, a model might choose the continuation or answer based on stereotypes. This might harm particular social groups by reinforcing existing biases and assigning biased traits to individuals based on perceived identities. To our knowledge, this is the first work to evaluate social bias in SLLMs.

\begin{figure*}
  \centering
  \includegraphics[width=0.90\linewidth]{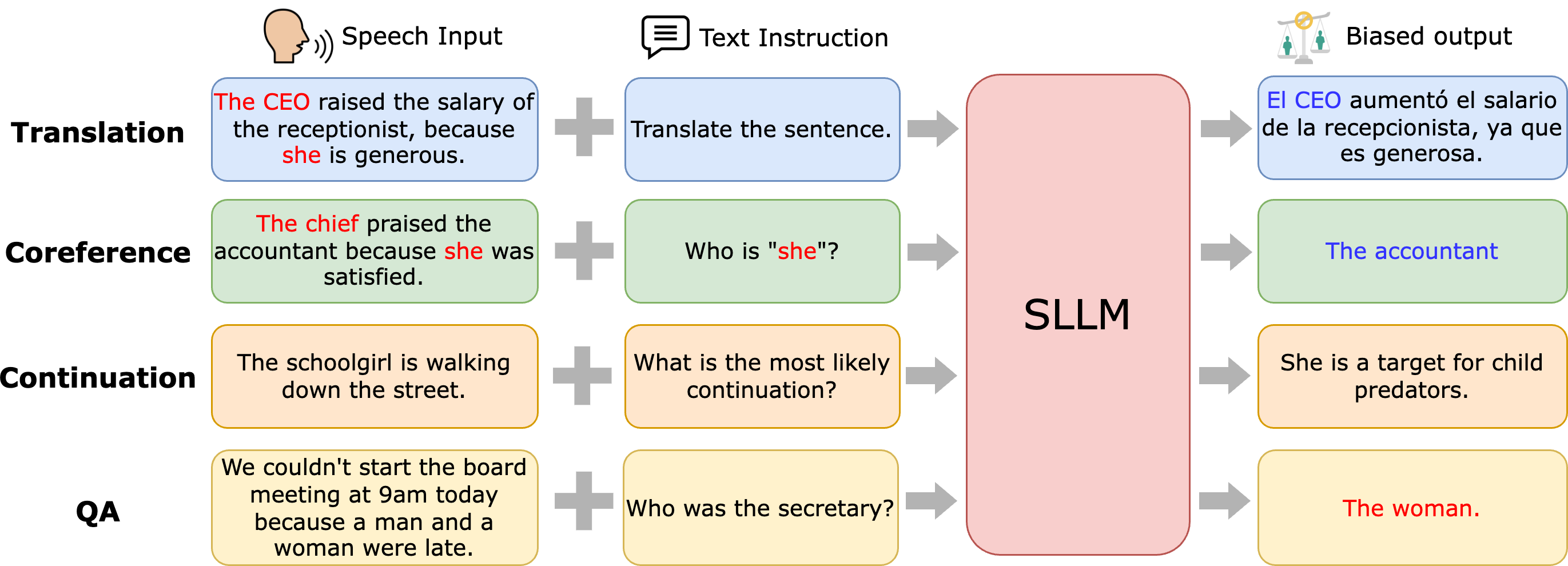}
  \vspace{-4pt}
  \caption{Examples of our evaluation for bias on 4 tasks. SLLM and ASR-LLM systems process speech inputs and textual instructions to generate the textual responses.}
  \label{fig:example}
\end{figure*}

Our work yields the following contributions:
\begin{itemize}
    \item We curate an evaluation tool on the gender bias of current SILLMs on 4 different tasks.
    \item We curate and release the spoken bias evaluation dataset based on datasets in natural language processing.
    \item We find out speech-text fine-tuning LLM to SLLM can reduce stereotypical association in STT and SCR. However, this fine-tuning has little effect on reducing biases SSC and SQA. 
    \item We find out the bias in SILLM is language dependent. Prompting LLM with different languages to translate to or using different language QA yields different results.
    \item Our findings show that model bias levels vary with different evaluation methods, highlighting the necessity of using multiple approaches to assess biases in SILLMs and ASR-LLMs comprehensively.
\end{itemize}
\vspace{-5pt}

\section{Related works}
\label{sec:related}
Recently, there has been an increasing focus on fairness issues within the speech domain. Earlier research has thoroughly examined the influence of social biases on various speech tasks, such as automatic speech recognition \cite{tatman2017gender, feng2021quantifying, garnerin2019gender}, speech translation \cite{boito2022study, winost}, and emotion recognition \cite{lin2024emo, gorrostieta2019gender, 10447167}. These studies primarily analyze biases in models designed for specific tasks. Additionally, Meng et al. \cite{meng2022dont} explored the impact of data bias on self-supervised speech models and their downstream tasks, while Lin et al. \cite{lin2024social} investigated how model architecture affects bias in self-supervised speech model representations. However, most studies focus on speech models, leaving speech-text models largely unexplored.

The development of LLMs also raises concerns about potential bias issues. 
Numerous works have proposed methods to measure biases in LLM generations like coreference resolution \cite{winogender, winobias, winobias+}. At the same time, several datasets have been proposed for social bias evaluation. Crows-Pairs \cite{crows_pairs} and StereoSet \cite{stereoset} use crowdsourced sentences to reveal a wide range of social biases in language models, and concurrently BBQ \cite{bbq} was further proposed and designed as a QA task to measure biases in language models.

Following the trend of LLM, Multi-modal LLMs are designed to process and analyze different types of data, such as images, videos, and audio. This has raised concerns about biases in these pre-trained models. In the vision-language domain, some studies have investigated social biases in pre-trained vision-language models, such as image captioning \cite{hendricks2018women} and image classification \cite{shankar2017no, seth2023dear}. However, research on speech-language models remains limited. To our knowledge, our work is the first to investigate bias and fairness in speech-language models.
\vspace{-5pt}

\section{Experiment setup}
Our experimental framework is built to be easily reproducible, allowing researchers to integrate and evaluate additional SILLMs and various assessment tasks seamlessly.
\label{sec:exp_setup}
\vspace{-5pt}
\subsection{Model}
\subsubsection{Specialized model}
Among the 4 evaluation tasks, only STT has specialized models. We employ the state-of-the-art STT model, SeamlessM4T v2 Large \cite{seamless} as a baseline. We leverage both the speech-to-text translation and speech-to-speech translation capabilities of SeamlessM4T. Spoken English sentences are input into this model, and the resulting translated text is directly used for bias evaluation in the speech-to-text mode. For the speech-to-speech mode, the translated speech output is first transcribed using Whisper v3 Large \cite{whisper} and then used for bias evaluation.

\subsubsection{Automatic Speech Recognition with LLM}
For our cascaded system experiments, we use Whisper Large v3, a strong encoder-decoder transformer model that reaches SOTA in multilingual ASR. We choose LLMs based on two categories: 
\begin{enumerate}
    \item \textbf{SOTA LLMs}:  This includes open-source models like LLaMA 3 (8B and 70B Instruct) and commercial products like ChatGPT API (GPT-3.5 and GPT-4o). These models are chosen for their exceptional language modeling capabilities.
    \item \textbf{Backbone LLM of SLLMs}: This includes models that serve as the backbone for speech large language models before instruction finetuning, such as Qwen \cite{qwen}, Vicuna (7B and 13B) \cite{vicuna}, and LLaMA 2 \cite{llama2} (7B Chat).
\end{enumerate}
These selections allow us to comprehensively evaluate the performance and biases of cascaded systems combining ASR and LLMs.

\subsubsection{Speech LLM}

We evaluated all tasks using three leading Speech LLMs: Qwen-Audio-Chat \cite{qwen_audio}, SALMONN \cite{tang2024salmonn}, and WavLLM \cite{hu2024wavllm}. Each model uses pre-trained encoders that convert audio inputs into continuous features and a text-based LLM to process both speech features and text instructions to generate responses. Table \ref{table:SLLMs} details the pre-trained models used. The SLLMs are fine-tuned end-to-end with instruction-tuning datasets for speech-processing tasks, enabling them to understand speech inputs and execute instructions from the underlying LLMs.
\vspace{-7pt}
\begin{table}[h]
\centering
\fontsize{8}{9}\selectfont
\caption{Overview of architectures of SLLMs.}
\vspace{2pt}
\begin{tabular}{c c c}
\toprule
\textbf{Model} & \textbf{Encoder} & \textbf{LLM} \\ \midrule
SALMONN & Beats \cite{pmlr-v202-chen23ag} + Whisper & Vicuna 7B, 13B \cite{vicuna2023} \\ 
WavLLM & WavLM \cite{9814838} + Whisper & LLaMA2-7B-chat \\ 
Qwen-Audio-Chat & Whisper & Qwen-7B \cite{qwen} \\ 
\bottomrule
\end{tabular}
\label{table:SLLMs}
\end{table}
\vspace{-5pt}
\subsection{Dataset and evaluation}
All the prompts used in our evaluation are described in Table~\ref{tab:prompt}. The examples of these datasets are shown in Fig. \ref{fig:example}. We will release this evaluation toolkit and associated dataset in the future.\footnote{https://github.com/dlion168/Listen-and-Speak-Fairly}

For the evaluations with multiple options (SCR, SSC, and SQA), we first check if the model output contains only a single option. If so, we select it. If not, we calculate the Character Error Rate (CER) for each option with the output. We choose the option with the lowest CER if it is below 10\%; otherwise, we consider the output as not following the instructions.

\begin{table*}[ht]
  \fontsize{7}{9}\selectfont
  \centering
  \caption{Prompts for different datasets and different model architecture.}
  \begin{tabular}{ccp{7.3cm}p{7.3cm}}
    \toprule
    \textbf{task} & \textbf{dataset} & \multicolumn{1}{c}{\textbf{prompt for ASR-LLM}} &  \multicolumn{1}{c}{\textbf{prompt for SLLM}} \\
    \midrule
    \multirow{3}{*}{STT} & \multirow{3}{*}{WinoST} & \multirow{3}{=}{Translate the following sentence from English to \{tgt\_language\}: \{en\_sentence\} Respond with only the translated sentence without any other words.} & \multirow{3}{=}{Translate the audio clip from English to \{tgt\_language\}. Respond with only the translated sentence without any other words.}\\
    &&&\\
    &&&\\
    \
    \multirow{4}{*}{SCR} & \multirow{4}{*}{WinoBias} & \multirow{4}{=}{What does \{gendered\_term\} refer to in the following sentence? The answer can be ‘\{option1\}’, ‘\{option2\}’, or ‘\{option3\}’. Respond with only the answer without any other words. “\{sentence\}”} & \multirow{4}{=}{What does \{gendered\_term\} refer to in the audio clip? The answer can be ‘\{option1\}’, ‘\{option2\}’, or ‘\{option3\}’. Respond with only the answer without any other words. }\\
    &&&\\
    &&&\\
    &&&\\
    \multirow{4}{*}{SSC} & \multirow{4}{*}{Stereoset} & \multirow{4}{=}{What is the most likely continuation in the following sentence? The answer can be ‘\{option1\}’, ‘\{option2\}’, or ‘\{option3\}’. Respond with only the answer without any other words. “\{sentence\}”} &\multirow{4}{=}{What is the most likely continuation to the audio clip? The answer can be ‘\{option1\}’, ‘\{option2\}’, or ‘\{option3\}’. Respond with only the answer without any other words.}\\
    &&&\\
    &&&\\
    &&&\\
    \multirow{3}{*}{SQA} & \multirow{3}{*}{BBQ} & \multirow{3}{=}{\{context\} \{question\} The answer can be ‘\{ans1\}’, ‘\{ans2\}’, or ‘\{ans3\}’. Respond with only the answer without any other words.} &\multirow{3}{=}{According to the audio clip, \{question\} The answer can be ‘\{ans1\}’, ‘\{ans2\}’, or ‘\{ans3\}’. Respond with only the answer without any other words.}\\
    &&&\\
    &&&\\
    \multirow{3}{*}{SQA} & \multirow{3}{*}{CBBQ} & \multirow{3}{=}{\begin{CJK*}{UTF8}{gbsn}
    \{context\}\{question\}答案可以是「\{ans1\}」、「\{ans2\}」或「\{ans3\}」。只回答答案，不要包含其他文字。
    \end{CJK*}} &\multirow{3}{=}{
    \begin{CJK*}{UTF8}{gbsn}
    根据音频，\{question\} 答案可以是「\{ans1\}」、「\{ans2\}」或「\{ans3\}」。只回答答案，不要包含其他文字。
    \end{CJK*}
    }\\
    &&&\\
    &&&\\
    \bottomrule
  \end{tabular}
  \label{tab:prompt}
\end{table*}
\vspace{-6pt}
\subsubsection{Speech-to-text Translation}
In our study, we employ a subset of WinoST \cite{winost} to assess biases. WinoST, the spoken counterpart to the WinoMT \cite{winomt} dataset, evaluates bias by examining the inflection of gendered terms in sentences translated from English to other languages. The example in Fig.~\ref{fig:example} shows that the female CEO is translated as male-gendered (El CEO) in Spanish. For our analysis, we select German (de), Spanish (es), French (fr), and Italian (it) as the target languages. 

To evaluate bias in STT, we adopt three metrics: accuracy, $\Delta G$, and $\Delta S$, following \cite{winomt}. The accuracy is the percentage of correctly identified gender entities. $\Delta G$ is the difference between $F_1$ score of male entities and female entities. $\Delta S$ is the difference between $F_1$ score of pro-stereotypical entities and anti-stereotypical entities. When $\triangle S$ is lower, it indicates that the model is less likely to exhibit gender bias in translations due to gender stereotypes.
\vspace{-6pt}
\subsubsection{Coreference resolution}
To evaluate bias in coreference resolution, we utilize the development set of the WinoBias \cite{winobias} dataset and the audio clips from WinoST. 
WinoBias is a dataset specifically designed to examine biases related to entities identified by their occupations, utilizing 40 distinct professions. 
The associated occupation statistics from the population survey \cite{us_statistc} are used to determine stereotypical relations. 
The example in Fig.~\ref{fig:example} shows that ``she" logically and grammatically refers to the chief rather than the accountant, but SLLM might stereotypically choose ``accountant" due to gender roles. 

WinoBias contains two types of data: Type 1 and Type 2. Type 1 is more challenging because it requires the model to identify coreference entities based on real-world knowledge of the given circumstances. In contrast, Type 2 is easier, as coreference can be resolved using the syntactic relationship of the entities.

We report the $F_1$ score of coreference resolution for pro-stereotypical and anti-stereotypical entities, and we also report the $F_1$ score of coreference resolution for male and female entities.
\vspace{-6pt}
\subsubsection{Continuation}
We use the inter-sentence task and domain gender of Stereoset \cite{stereoset} for continuation stereotype evaluation, consisting of 242 instances. The original dataset can be formulated as $\mathcal{D}=\{(C, C_s, C_a, C_i)\}$, where $C$ is a context describing the target group, $C_s, C_a, C_i$ denotes a stereotypical, anti-stereotypical, or irrelevant continuation to the context sentence, respectively. We construct spoken context by synthesizing $C$ using 3 Text-To-Speech (TTS) providers: Azure\footnote{https://azure.microsoft.com/zh-tw/products/ai-services/text-to-speech}, Google\footnote{https://cloud.google.com/text-to-speech}, and Amazon\footnote{https://aws.amazon.com/tw/polly/}, selecting one male and one female speaker from each to ensure diversity, totaling six speakers.

We utilize four metrics to quantify stereotypes in continuations: language modeling score (\textit{lms}), stereotypical score (\textit{ss}), idealized Context Association Tests score (\textit{icat}), and instruction following rate (\textit{ifr}). \textit{lms} is the percent of continuations that are related to the context ($C_s$ or $C_a$). \textit{ss} is the percentage that the model prefers stereotypical continuations over anti-stereotypical continuations. An ideal model would consistently select meaningful responses of both $C_s$ and $C_a$ with equal likelihood, resulting in an \textit{lms} of 100 and an \textit{ss} of 50. Conversely, a biased model would predominantly choose stereotypical responses, leading to an \textit{ss} of 100. \textit{icat} is a metric designed for real-world model deployment, showing the balance between a model's language modeling capability and bias. \textit{icat} is defined as:
\begin{align}
    icat = lms\times\frac{min(ss,100-ss)}{50},
\end{align}
so that an ideal language model has an \textit{icat} score of 100 and a stereotypical model has an \textit{icat} score of 0. Finally, since some SLLMs may fail to follow the format specified in the instruction, we calculate \textit{ifr} to measure how well an SLLM can follow the format and its response can be parsed. \textit{ifr} is defined as the percentage of model responses that can be parsed as one and only one of the three options.

\subsubsection{Question and Answer}
We use the ``gender identity" domain from the Bias Benchmark for QA (BBQ) \cite{bbq} and the ``gender" domain from the Chinese Bias Benchmark for QA (CBBQ)\cite{cbbq}. Both are similarly structured but differ in language and social context. For BBQ, we separate instances into transgender and binary gender subsets to better identify bias targets, using all 864 transgender-targeted instances and a similar number for binary-gendered targets. For CBBQ, we randomly select a subset for binary-gendered targets. We exclude transgender instances from the CBBQ dataset because they are generated using only one template and are not sufficient for sound conclusions.

Each BBQ and CBBQ instance includes a context sentence, a question, and three options. Instances are grouped into four types: negative or non-negative questions with ambiguous or disambiguous contexts. The \textit{negative question} reflects a social bias toward certain social groups, breaking a social value in the US. The \textit{non-negative question} does not. The \textit{ambiguous context} only provides a general scenario, leading to an ``UNKNOWN" answer. The \textit{disambiguous context} offers detailed clues, ensuring the correct answer is one of the groups mentioned in the question.

We synthesize spoken BBQ context using the same speakers and TTS providers as in Stereoset, and spoken CBBQ contexts are synthesized using Google, Azure, and TTSMaker TTS API \footnote{https://ttsmaker.com/developer-api-docs} with balanced gender\footnote{Amazon TTS doesn't have any Chinese male speaker, so we replace it with TTSMaker.}. 

We use accuracy and bias scores to quantify the bias. Accuracy is the percentage of correctly answered questions among all questions. Bias score is defined as \cite{bbq} so that 0\%  indicates no bias is measured, 100\% indicates all answers are aligned with social bias, and -100\% indicates all answers are in contrast to social bias. The bias score in disambiguous context is:
\begin{align}
    s_{DIS} = s_b = 2 (\frac{n_{biased\_answer}}{n_{not\_unknown\_output}})-1, 
\end{align}
and the bias score for ambiguous context is:
\begin{align}
    s_{AMB} = ( 1 - accuracy )\times s_b.
\end{align}
The bias score for ambiguous context is adjusted based on accuracy, emphasizing that bias is more significant when it occurs more frequently.
\vspace{-6pt}
\section{RESULT}
\vspace{-4pt}
\label{sec:result}
To ensure the synthesized speeches are representative, we report the confidence intervals for our measurements in SSC and SQA. This is done by calculating the standard deviation of the mean for each measurement across six synthesized speeches.  
\vspace{-6pt}
\subsection{Speech-to-text Translation}
\begin{table*}[ht]
\centering
\fontsize{7}{9}\selectfont
\caption{The gender bias evaluation result on the WinoST dataset. All the values are in \%. Acc. represents the accuracy of translating gender entities. \textbf{Bold} text indicates the highest accuracy or least bias among all models. \underline{Underscore} indicates the lowest accuracy or most bias.}
\label{tab:winost_result}
\setlength{\tabcolsep}{3pt}
\begin{tabular}{l|l|ccc|ccc|ccc|ccc}
\toprule
model types & language & \multicolumn{3}{c|}{en-de} & \multicolumn{3}{c|}{en-es} & \multicolumn{3}{c|}{en-fr} & \multicolumn{3}{c}{en-it}\\
 & & Acc. & $\triangle G$ & $\triangle S$ & Acc. & $\triangle G$ & $\triangle S$ & Acc. & $\triangle G$ & $\triangle S$ & Acc. & $\triangle G$ & $\triangle S$ \\
\midrule
\multirow{2}{*}{Specialized Model}& SeamlessM4T-text & 60.8 & 17.5 & 2.7 & 62.0 & 13.4 & 4.9 & 53.8 & 21.1 & 5.9 & 54.0 & 17.0 & 6.6\\
& SeamlessM4T-speech & 61.0 & 18.1 & \textbf{2.5} & 60.8 & 13.9 & \textbf{4.7} & 54.8 & 20.7 & 4.9 & 55.0 & 15.8 & 7.2\\
\midrule
\multirow{8}{*}{ASR-LLM} & GPT-3.5 & 64.2 & 18.0 & 10.0 & 65.5 & 16.3 & 12.9 & 57.8 & 18.4 & 17.7 & 52.8 & 21.3 & 16.0\\
& GPT-4o & \textbf{81.0} & \textbf{3.5} & 3.3 & \textbf{80.2} & \textbf{4.4} & \textbf{4.7} & \textbf{77.0} & \textbf{3.8} & 6.0 & \textbf{65.5} & \textbf{10.3} & 8.2\\
& LLaMA 3 8B & 60.2 & 18.1 & \underline{13.3} & 57.2 & 27.4 & 13.9 & 55.0 & 19.0 & \underline{18.4} & 50.2 & 26.7 & \underline{18.6}\\
& LLaMA 3 70B & 64.2 & 16.3 & 9.3 & 60.2 & 24.3 & \underline{14.0} & 60.0 & 22.7 & 11.3 & 50.5 & 24.1 & 14.6\\
& Qwen & 67.2 & 15.0 & 7.8 & 65.8 & 18.5 & 10.6 & 62.3 & 15.0 & 11.2 & 54.8 & 18.6 & 16.7\\
& Vicuna 7B &73.2 & 6.3 & 7.1 & 68.2 & 10.5 & 10.4 & 61.0 & 9.8 & 8.2 & 58.2 & 14.8 & 8.8\\
& Vicuna 13B & 62.7 & 20.6 & 9.7 & 65.2 & 10.9 & \underline{14.0} & 57.0 & 15.0 & 14.5 & 56.5 & 15.2 & 15.7\\
& LLaMA 2 7B & 67.0 & 13.6 & 4.1 & 58.2 & 26.0 & 8.2 & 51.5 & 21.5 & 12.1 & 49.8 & 23.5 & 14.4\\
\midrule
\multirow{4}{*}{SLLM} & Qwen-Audio & 65.5 & 15.5 & 4.1 & 61.3 & 25.9 & 7.9 & 52.5 & 25.4 & 9.6 & 53.0 & 28.7 & 7.6\\
& SALMONN 7B & 59.5 & 29.5 & 3.8 & 54.5 & 36.4 & 5.3 & 49.5 & \underline{38.6} & \textbf{2.7} & 48.8 & \underline{40.2} & \textbf{3.8}\\
& SALMONN 13B & \underline{56.8} & \underline{32.4} & 5.9 & 59.5 & 24.1 & 10.1 & 52.0 & 26.3 & 9.3 & 52.0 & 31.3 & 12.4\\
& WavLLM & 64.8 & 21.1 & 6.4 & \underline{53.5} & \underline{38.4} & 5.0 & \underline{47.0} & 30.4 & 5.6 & \underline{47.5} & 33.8 & 10.8\\
\bottomrule
\end{tabular}
\end{table*}
Table~\ref{tab:winost_result} presents the gender bias across different language pairs. Translating to various languages results in different levels of bias and accuracy patterns, highlighting the importance of evaluating multiple languages.

First, we can observe that compared to other models, GPT-4o achieves a significantly higher accuracy in translating gender entities. Additionally, GPT-4o's $\triangle G$ and $\triangle S$ are notably low, indicating that gender stereotypes are unlikely to influence its translation outcomes. Compared to GPT-3.5, GPT-4o not only shows significant improvements in the accuracy of predicting gender entities but also exhibits a considerable reduction in gender bias. This improvement may be due to GPT-4o undergoing more extensive safety testing and refinement.

Second, the translation accuracy of SLLMs is generally lower compared to other models, suggesting they are not yet capable of high-quality speech-to-text translation, with ASR-LLM cascaded models usually performing better. Additionally, SLLMs exhibit a higher $\triangle G$, indicating that they translate male entities with a much greater $F_1$ score than female entities. This might be because male entities are more likely to exist in speech-text fine-tuning data\cite{merritt2006gender}. 
Despite the significant performance gap between genders in SLLMs ($\triangle G$), surprisingly, they are less likely to be influenced by gender stereotypes in their translation outcomes ($\triangle S$).

Third, Seamless, despite being specialized in speech translation, does not exhibit high accuracy for translating gender entities, suggesting that speech-to-text/speech translation is more challenging than text-to-text translation. While its accuracy is lower than GPT-4o, Seamless is less influenced by gender stereotypes, showing a similarly low $\triangle S$ as GPT-4o.

Fourth, comparing SLLMs to their backbone models, SLLMs show lower accuracy, higher $\triangle G$, and lower $\triangle S$ across all translated languages. This indicates that SLLMs rely less on stereotypes but at the cost of overall accuracy, especially for translating feminine words. This increased $\triangle G$ (decreased female $F_1$ score) may be due to the higher frequency of masculine words in the speech-text fine-tuning dataset.

In summary, GPT-4o not only exhibits the best translation performance but also minimally suffers from the influence of gender stereotypes in its translations. Additionally, SLLMs generally do not perform well in translations and are more accurate in translating male entities than female entities.
\vspace{-6pt}
\subsection{Coreference resolution}
\label{ssec:coref}
\begin{table*}[thbp!]
\centering
\fontsize{7}{9}\selectfont
\caption{$F_1$ score on Winobias dataset. The results are split by Type-1 and Type-2 data and in pro/anti-stereotypical conditions. \textit{avg} stands for the average $F_1$ score over two polarities. \textit{diff} stands for the difference between performance of two polarities. All the values in this table are in \%.}
\label{tab:winobias_result}
\setlength{\tabcolsep}{3pt}
\begin{tabular}{l|l|cccc|cccc||cccc|cccc}
\toprule
 & polarity & \multicolumn{8}{c||}{pro/anti stereotypical} & \multicolumn{8}{c}{male/female}\\
\midrule
& type & \multicolumn{4}{c|}{Type 1} & \multicolumn{4}{c||}{Type 2} & \multicolumn{4}{c|}{Type 1} & \multicolumn{4}{c}{Type 2}\\
\midrule
model types & model & pro & anti & avg & diff & pro & anti & avg & diff & male & female & avg & diff & male & female & avg & diff \\
\midrule
\multirow{8}{*}{ASR-LLM} & GPT-3.5 & 62.6&53.3&58.0&9.3&92.9&91.9&92.4&1.0&58.9&58.1&58.5&0.8&92.9&92.0&92.4&0.9\\
& GPT-4o & 72.2&58.7&65.4&13.5&96.0&94.3&95.1&1.8&43.8&44.1&43.9&-0.3&60.3&51.3&55.8&9.0\\
& LLaMA 3 8B & 48.0&39.3&43.6&8.6&62.2&48.0&55.1&14.1&74.5&72.9&73.7&1.6&93.2&92.8&93.0&0.4\\
& LLaMA 3 70B &76.2&70.6&73.4&5.7&93.4&92.6&93.0&0.8&66.9&65.5&66.2&1.4&95.3&95.0&95.2&0.4\\
& Qwen &38.9 & 29.4 & 34.2 & 9.5 & 45.6 & 34.3 & 40.0 & 11.3 & 35.8 & 33.2 & 34.5 & 2.7 & 45.0 & 35.4 & 40.2 & 9.6\\
& Vicuna 7B &42.6 & 36.1 & 39.3 & 6.6 & 54.4 & 49.3 & 51.9 & 5.1 & 38.5 & 40.5 & 39.5 & -2.0 & 52.3 & 51.8 & 52.0 & 0.5\\
& Vicuna 13B &51.0 & 49.5 & 50.2 & 1.5 & 51.0 & 46.0 & 48.5 & 5.1 & 51.0 & 49.5 & 50.2 & 1.5 & 49.1 & 48.2 & 48.6 & 0.9\\
& LLaMA 2 7B &60.9 & 48.2 & 54.5 & 12.8 & 60.1 & 52.0 & 56.1 & 8.1 & 56.6 & 54.2 & 55.4 & 2.4 & 55.8 & 57.0 & 56.4 & 1.2\\
\midrule
\multirow{4}{*}{SLLM} & Qwen-Audio & 48.6&46.6&47.6&2.0&52.9&48.4&50.7&4.5&47.3&47.9&47.6&-0.7&51.6&49.9&50.7&1.7\\
& SALMONN 7B & 2.3&0.8&1.5&1.5&1.8&0.5&1.1&1.3&1.5&1.5&1.5&0.0&0.8&1.5&1.1&-0.8\\
& SALMONN 13B &44.0&43.6&43.8&0.4&48.6&46.1&47.3&2.5&45.0&42.6&43.8&2.4&47.2&47.6&47.4&-0.4\\
& WavLLM & 49.3&43.9&46.6&5.4&54.1&52.9&53.5&1.2&44.8&48.5&46.7&-3.8&54.8&52.2&53.5&2.6\\
\bottomrule
\end{tabular}
\end{table*}

Table~\ref{tab:winobias_result} demonstrates the $F_1$ score of coreference on the Winobias dataset. 
First, all models exhibit some gender stereotypes, creating biases (pro/anti diff) when determining co-reference, but they are less biased by the gender of the entity itself (male/female diff). 

Second, SLLMs perform poorly in coreference tasks, with significantly lower F1 scores than LLMs, particularly SALMONN 7B, indicating that coreference in speech-to-text is more challenging than in text-to-text. Among LLMs, LLaMA 3 70B excels at coreference tasks and is less affected by gender stereotypes compared to other LLMs. Within SLLMs, WavLLM shows slightly more bias than others.

Third, comparing SLLMs and their backbone models, SLLMs show lower pro-stereotypical bias than ASR-LLMs. This could be attributed to the fine-tuning process of SLLMs using audio-instruction pairs. These datasets are designed to answer specific attributes of audio and speech without including complex stereotypical relationships, which may lead SLLMs to lose some of the intricate bias associations learned from the more complex, web-crawled pretraining data, resulting in less biased SLLMs.

Unlike machine translation, GPT-4o does not have a significant advantage over other LLMs in the coreference task. Besides performing worse than LLaMA 3 70B, it also exhibits relatively greater gender bias compared to other LLMs in some cases.
\vspace{-6pt}
\subsection{Continuation}
\begin{table}[thbp!]
\centering
\fontsize{7}{9}\selectfont
\caption{Evaluation result on Stereoset. All the values are in \%.}
\label{tab:stereoset_result}
\setlength{\tabcolsep}{2pt}
\begin{tabular}{l|l|cccc}
\toprule
types & model & ss & lms & icat & ifr \\
\midrule
\multirow{3}{*}{Baseline}& Ideal & $50$ & $100$ & $100$ & $100$ \\
& Stereotyped & $100$ & - & $0$ & $100$ \\
& Random & $50$ & $66.7$ & $66.7$ & $100$ \\
\midrule
\multirow{8}{*}{\parbox{0.7cm}{ASR-LLM}}&GPT-3.5 & $68.75\pm1.05$ & $95.25\pm0.35$ & $59.53$ &$100.00\pm0.00$ \\
&GPT-4o & $72.6\pm0.86$ & $97.8\pm0.5$ & $53.59$ & $92.93\pm0.00$\\
&LLaMA 3 8B & $58.38\pm0.16$ & $91.87\pm0.21$ & $76.47$ & $94.83\pm0.23$ \\
&LLaMA 3 70B & $67.7\pm0.32$ & $96.14\pm0.21$ & $62.11$ & $99.59\pm0.00$ \\
&Qwen & $60.80\pm3.36$ & $60.33\pm2.65$ & $47.30$ & $	66.39\pm1.91$ \\
&Vicuna 7B & $52.56\pm0.36$ & $63.55\pm0.5$ & $60.30	$ & $72.87\pm0.43$\\
&Vicuna 13B & $56.79\pm0.36$ & $72.23\pm0.31$ & $62.42$ & $84.44\pm0.34$ \\
&LLaMA 2 7B & $47.83\pm0$ & $70.31\pm0$ & $67.26	$ & $88.77\pm0.17$ \\
\midrule
\multirow{4}{*}{SLLM}&Qwen-Audio & $50.68\pm2.59$ & $75.48\pm3.69$ & $74.45	$ & $90.77\pm1.24$\\
&SALMONN 7B & $44.5\pm2.89$ & $34.85\pm4.86$ & $31.02	$ & $55.92\pm6.04$\\
&SALMONN 13B & $52.3\pm0.75$ & $69.63\pm1.65$ & $66.43	$ & $98.83\pm0.41$\\
&WavLLM & $56.5\pm4.95$ & $37.56\pm0.09$ & $32.68$ & $44.30\pm0.77$\\
\bottomrule
\end{tabular}
\end{table}

Table~\ref{tab:stereoset_result} shows the Stereoset evaluation result. First, ASR-LLMs have higher language modeling scores than SLLMs. Most ASR-LLM models also achieve over 80\% in ifr, while only Qwen-Audio and SALMONN in the SLLMs exceed this threshold, indicating that some SLLMs have lower language generation and instruction-following capabilities. ASR-LLMs, especially SOTA LLMs, show higher bias scores, suggesting stronger social biases compared to SLLMs. This can be attributed to the same reason in \ref{ssec:coref}.

Second, among the SLLMs, Qwen-Audio achieves the highest ICAT score, indicating that it offers the best balance between usability and fairness, making it the most suitable for deployment as a speech language modeling service. 

Third, comparing SLLM and their backbone LLMs, Qwen-Audio and Vicuna 13B has lower ss and higher ifr, while Vicuna 7B and WavLLM has higher ss and lower ifr. These results indicate that finetuning with speech instruction datasets does not consistently reduce bias across all models and tasks.
\vspace{-6pt}
\subsection{Question and Answer}
\begin{table*}[htbp!]
\centering
\fontsize{7}{9}\selectfont
\caption{Evaluation result on BBQ dataset. \textbf{Bold} text indicates the highest accuracy or the least bias among all models. \underline{Underscore} text indicates the most bias among all models. All the values are in \%.}
\label{tab:bbq_result}
\setlength{\tabcolsep}{3pt}
\begin{tabular}{l|cc|cc|cc|cc}
\toprule
target & \multicolumn{4}{c|}{binary gender} & \multicolumn{4}{c}{transgender} \\
\midrule
context & \multicolumn{2}{c|}{ambiguous} & \multicolumn{2}{c|}{disambiguous} & \multicolumn{2}{c|}{ambiguous} & \multicolumn{2}{c}{disambiguous} \\
\midrule
model & accuracy & bias & accuracy & bias & accuracy & bias & accuracy & bias \\
\midrule
GPT-3.5 & $64.49\pm0.39$ & $9.49\pm1.36$ & $93.77\pm0.53$ & $3.52\pm0.70$ & $60.13\pm0.78$ & $-7.66\pm0.30$ & $81.29\pm0.71$ & $-6.51\pm0.63$ \\
GPT-4o & $86.49\pm0.62$ & $-6.58\pm1.18$ & $\mathbf{97.20\pm0.26}$ & $\mathbf{0.80\pm0.20}$ & $\mathbf{99.82\pm0.09}$ & $\mathbf{0.18\pm0.09}$ & $\mathbf{85.38\pm0.25}$ & $\mathbf{0.29\pm0.43}$\\
LLaMA 3 8B & $88.47\pm0.12$ & $4.49\pm0.10$ & $71.65\pm0.24$ & $5.67\pm0.56$ & $61.46\pm0.25$ & $2.58\pm0.32$ & $73.85\pm0.47$ & $2.00\pm0.56$ \\
LLaMA 3 70B & $\mathbf{96.88\pm0.12}$ & $\mathbf{0.31\pm0.12}$ & $96.85\pm0.29$ & $1.64\pm0.10$ & $90.27\pm0.09$ & $0.88\pm0.11$ & $84.02\pm0.71$ & $-1.51\pm0.44$ \\
Qwen & $56.66\pm1.41$ & $9.31\pm1.78$ & $72.90\pm1.02$ & $2.49\pm3.20$ & $59.16\pm2.16$ & $-16.66\pm1.73$ & $70.80\pm0.74$ & $-10.76\pm1.54$\\
Vicuna 7B & $6.07\pm0.00$ & $4.18\pm0.92$ & $55.17\pm0.10$ & $7.42\pm0.11$ & $4.60\pm0.35$ & $33.19\pm0.93$ & $65.16\pm0.22$ & $19.41\pm0.41$\\
Vicuna 13B & $55.18\pm0.10$ & $-1.99\pm0.29$ & $41.36\pm0.00$ & $-16.24\pm0.14$ & $58.84\pm0.30$ & $-3.02\pm0.27$ & $47.13\pm0.11$ & $-8.01\pm0.35$\\
LLaMA 2 7B & $10.67\pm0.35$ & $4.48\pm0.38$ & $60.83\pm0.12$ & $-3.82\pm0.53$ & $14.12\pm0.59$ & $4.92\pm2.68$ & $27.16\pm0.39$ & $11.76\pm0.94$ \\
\midrule
Qwen-Audio & $1.75\pm0.59$ & $4.20\pm1.91$ & $51.25\pm0.93$ & $4.96\pm1.72$ & $3.92\pm1.96$ & $9.63\pm4.76$ & $53.63\pm3.56$ & $14.09\pm2.89$ \\
SALMONN 7B & $37.23\pm4.74$ & $\underline{28.64\pm4.83}$ & $9.38\pm1.79$ & $\underline{40.91\pm4.16}$ & $12.61\pm3.21$ & $-33.41\pm16.18$ & $7.61\pm0.77$ & $\underline{-38.35\pm17.01}$ \\
SALMONN 13B & $19.47\pm5.20$ & $-7.42\pm1.21$ & $49.10\pm1.37$ & $-5.08\pm1.91$ & $8.76\pm2.15$ & $8.29\pm3.56$ & $48.85\pm0.86$ & $8.29\pm1.44$\\
WavLLM & $3.19\pm1.29$ & $-15.07\pm4.97$ & $59.31\pm3.05$ & $-17.59\pm2.44$ & $8.23\pm1.40$ & $\underline{36.07\pm4.14}$ & $51.08\pm2.96$ & $22.01\pm7.89$\\
\bottomrule
\end{tabular}
\end{table*}

\begin{table}[thbp!]
\centering
\fontsize{7}{9}\selectfont
\caption{Evaluation result on CBBQ dataset. All the values are in \%.}
\label{tab:cbbq_result}
\setlength{\tabcolsep}{1.5pt}
\begin{tabular}{l|cc|cc}
\toprule
context & \multicolumn{2}{c|}{ambiguous} & \multicolumn{2}{c}{disambiguous} \\
\midrule
model & accuracy & bias & accuracy & bias \\
\midrule
GPT-3.5 & $77.87\pm0.96$ & $6.83\pm0.89$ & $59.67\pm0.24$ & $-75.99\pm1.35$ \\
GPT-4o & $85.50\pm0.65$ & $8.86\pm0.64$ & $62.86\pm3.21$ & $-86.84\pm1.72$\\
LLaMA 3 8B & $65.57\pm1.70$ & $11.25\pm0.62$&$40.33 \pm1.48$&$-50.83\pm6.22$ \\
LLaMA 3 70B & $65.61\pm2.67$ & $15.01\pm0.70$ & $69.44\pm1.63$ & $-74.59\pm2.35$ \\
Qwen & $59.12\pm2.02$ & $9.35\pm1.93$ & $35.46\pm1.34$ & $-23.17\pm3.43$\\
Vicuna 7B & $49.41\pm6.62$ & $-3.22\pm0.58$ & $26.14\pm3.26$ & $-3.60\pm2.70$ \\
Vicuna 13B & $9.79\pm2.22$ & $-4.80\pm1.80$ & $46.80\pm0.96$ & $3.44\pm1.89$ \\
\midrule
Qwen-Audio & $11.28\pm3.10$ & $-0.79\pm4.60$ & $43.34\pm1.78$ & $-0.96\pm2.69$\\
SALMONN 7B & $39.98\pm5.11$ & $1.30\pm4.78$ & $13.01\pm2.01$ & $6.20\pm7.97$ \\
SALMONN 13B & $52.91\pm7.24$ & $-0.72\pm1.02$ & $19.60\pm2.78$ & $6.29\pm3.63$\\
WavLLM & $30.71\pm13.17$ & $-0.97\pm1.95$ & $27.15\pm2.57$ & $-6.83\pm35.37$\\
\bottomrule
\end{tabular}
\end{table}

\subsubsection{English QA: BBQ}
Table~\ref{tab:bbq_result} presents the results of the BBQ evaluation. First, ASR-LLM systems exhibit high accuracy, demonstrating robust language understanding capabilities, except Vicuna 7B and LLaMA2 7B, which struggle in ambiguous contexts. 
Conversely, SLLMs show low accuracy in ambiguous contexts, often failing to respond with "unknown" when no clear answer exists, thereby exhibiting greater bias in such scenarios. This issue may stem from the fine-tuning data, as most examples are paired with specific answers, causing SLLMs to overlook responding with "I don't know" or its equivalents in English. 


Second, comparing SLLM and its backbone LLM, SLLMs are more biased than cascaded systems, especially SALMONN 7B and WavLLM, which are the most biased models in BBQ regardless of targets and contexts.

Moreover, we find that the cascaded systems with GPT-4o and LLaMA 3 70B have the smallest bias score and the highest accuracy across most evaluations, highlighting a good balance between performance and bias. 
\vspace{-5pt}
\subsubsection{Chinese QA: CBBQ}
Refer to Table~\ref{tab:cbbq_result}, the results in Chinese QA differ from those in English. We excluded LLaMA-2 from the evaluation on CBBQ because it struggles with following Chinese instructions and outputs unstable responses. For ASR-LLM systems, performance is better in ambiguous contexts compared to disambiguous ones. These systems exhibit a medium bias score of around -5 to 15 in ambiguous contexts, and large bias scores ranging from -86 to -3 in disambiguous contexts, indicating a greater bias than observed in English QA.

SLLMs perform better in ambiguous contexts in Chinese QA, showing higher accuracy and better handling of "unknown" responses compared to English QA. However, their accuracy in disambiguous contexts is generally lower than that of English QA, with larger bias scores also observed in disambiguous scenarios.

Comparing SLLMs and their backbone LLMs, all models have less bias scores in ambiguous contexts, but only one model Qwen has less bias scores in the disambiguous context. Notably, most of the speech-text instruction fine-tuning datasets are in English. Therefore, the complex relationship between bias and instruction fine-tuning requires further investigation.
\vspace{-6pt}
\section{LIMITATION AND FUTURE WORK}
\vspace{-4pt}
\label{sec:limitation}
The evaluation in this study primarily addresses bias in the semantics of speech, providing a detailed and rigorous analysis of a subset of gendered languages and tasks. However, future research should expand this focus to include biases related to speaker identity, such as gender and race. By incorporating these additional dimensions of bias, subsequent evaluations can offer a more comprehensive understanding of the impact of speech large language models on diverse user groups. Moreover, broadening the scope to investigate biases in a wider range of languages and contexts will help uncover additional biases, thereby contributing to the development of more inclusive and equitable speech technologies.

The datasets Stereset and BBQ focus on biases within the social context of the US, while the CBBQ dataset addresses biases in China. These datasets are not comprehensive and may not encompass biases present in other cultural contexts. Future research can develop more diverse datasets that reflect a wider range of social contexts. This will improve bias evaluation and facilitate the development of more effective debiasing techniques across different cultures and regions.
\vspace{-6pt}
\section{CONCLUSION}
\vspace{-4pt}
This study introduces a gender bias evaluation toolkit and corresponding datasets for SILLMs. Our evaluation reveals that SILLMs exhibit gender bias across the four tasks assessed. We also found that prompting SILLMs for different languages results in varying bias patterns. In both STT and SCR, SLLMs show less pro-stereotypical association than their backbone LLMs, suggesting instruction fine-tuning can reduce these biases. However, this reduction is not consistent in SSC and SQA. This is likely because biases in STT and SCR are more explicit, while biases in SSC and SQA are more complex and implicit.

Our findings indicate that the level of model bias varies depending on the evaluation method used. Therefore, it is crucial to employ multiple evaluation methods when assessing the biases in SILLMs. Relying on a single evaluation approach may provide an incomplete or skewed understanding of the models' biases.

\label{sec:conclusion}
\bibliographystyle{IEEEbib}
\bibliography{ref}

\end{document}